\documentclass[twocolumn,showpacs,prl,10pt]{revtex4-1}
\usepackage[english]{babel}
\usepackage{amsmath,amssymb} 
\usepackage{times}
\usepackage{epsfig} 

\begin{document}

\title{Conductivity in a disordered one-dimensional system of interacting fermions}

\author{O. S. Bari\v si\' c$^{1}$ and P. Prelov\v sek$^{2,3}$}
\affiliation{$^1$Institute of Physics, HR-10000 Zagreb, Croatia}
\affiliation{$^2$J.\ Stefan Institute, SI-1000 Ljubljana, Slovenia}
\affiliation{$^3$ Faculty of Mathematics and Physics, University of
Ljubljana, SI-1000 Ljubljana, Slovenia}

\begin{abstract}

Dynamical conductivity in a disordered one-dimensional model of interacting 
fermions is studied numerically at high temperatures and in the weak-interaction
regime in order to find a signature of many-body localization and 
vanishing d.c. transport coefficients. On the contrary, we find in the regime of moderately 
strong local  disorder that the d.c. conductivity $\sigma_0$ scales linearly with the 
interaction strength while being exponentially 
dependent on the disorder. According to the behavior of the charge stiffness evaluated at the fixed number of particles, the absence of the many-body localization seems related to an increase of the effective localization length with the 
interaction.

\end{abstract}


\maketitle
The interplay of correlations and disorder in fermionic systems is one
of the challenging  open questions in the solid state physics. The 
phenomenon of Anderson localization of single-electron eigenstates\cite{ande} 
is by now  well understood in systems of noninteracting (NI) fermions. 
In particular, in one-dimensional (1D) systems all states become 
localized\cite{mott} for arbitrary small disorder\cite{abra} and hence 
there is no d.c. linear transport response at  any temperature $T \geq 0$.
However, it had been long ago realized\cite{flei} that correlations among 
electrons as introduced via Coulomb electron-electron repulsion 
could qualitatively change transport properties of the system.  

So far, firm results and conclusions have been reached for the 
$T=0$ ground state of 1D tight-binding fermionic system with a diagonal 
Anderson disorder. In particular, it has been shown by the 
density-matrix renormalization-group (DMRG)
numerical studies\cite{schm,urba} that in spite of correlations the 
many-body (MB)  states remain localized, preventing the d.c. transport.
The $T>0$ behavior appears to be much harder to deal with\cite{natt,gorn}
and, at present,  the existence of MB localization beyond the ground state is
controversial.\cite{daml} The most interesting conjecture emerging from an involved
analytical calculation\cite{bask} predicts a finite-temperature phase transition
between the MB insulator at $T<T^*$ and a conductor at $T>T^*$. 
Since such a transition in fact implies a qualitative change of character of MB
states across the eigenspectrum it as relevant and highly nontrivial to  
study systems at high $T \to \infty$.\cite{ogan} In this context, recent studies of energy-level statistics,\cite{ogan}
the effective hopping in the configuration space\cite{mont} and the decay 
of correlation functions\cite{pal} indicate a possible MB localization
at very large disorder strength $W$.\cite{mont} The conclusions from the scaling analysis of the conductivity of such 
models appear similar,\cite{berk} as well as the time evolution and the entanglement 
of wave-functions is concerned.\cite{zpp} 

On the other hand, recent direct numerical evaluation of the $T>0$ 
transport coefficients in  disordered anisotropic XXZ model\cite{kara} 
(model being equivalent in 1D to a tight-binding  fermionic  system
with nearest-neighbor interaction)
does not show any indication of a crossover to a MB localization at low
$T$ or at larger $W$. This  questions the conductor-insulator phase diagram 
and the relation to above mentioned studies.

Our aim is to extend previous numerical study\cite{kara} of transport
properties of the 1D disordered system, modeled by the $t$-$V$ model of
spinless fermions, in order to explore the phase
diagram at high $T$ with respect to the d.c. conductivity $\sigma_0$.
In contrast to most previous works, in which the interaction strength 
$\Delta=V/(2t) $ has been mainly kept fixed and possible MB localization 
has been considered at large disorder values $W$,  we  start with a disordered system 
of NI electrons ($\Delta=0$), characterized by the vanishing d.c. transport
at all $T$, i.e. $\sigma_0 =0$.  By increasing gradually, at fixed $W$, the repulsive interaction $\Delta>0$ we monitor a possible conductor-insulator transition in $\sigma_0$.  Dealing with a finite-size system, instead of a singular behavior we expect that the insulator-transition transition should manifest itself as a crossover in $\sigma_0$ vs. $\Delta$.
This crossover can be then used as a signature of a qualitative (gradual or abrupt) 
change of MB states with respect to the d.c. transport in the thermodynamic  
limit $T \to \infty$.  

As the prototype model for the interplay of correlations and disorder we study
the disordered 1D $t$-$V$ model. The Hamiltonian represents a tight-binding 
band of spinless fermions on a chain, the repulsion occurs between nearest 
neighbors, while the disorder is in site energies,
\begin{equation}
H= -t \sum_i( c^\dagger_{i+1} c_i  + {\rm h.c.}) + V\sum_i n_{i+1} n_i +
\sum_i \epsilon_i n_i~. \label{tv}
\end{equation}
By choosing site energies randomly in the 
interval $-W < \epsilon_i < W$, we obtain in the NI limit $V=0$ the
Anderson-localization model. In  order to avoid the interaction-induced Mott-type insulator at $\Delta=V/(2t) >1$, we restrict our study to the regime $\Delta<1 $ 
(note that for $\Delta =1$ the model (\ref{tv}) can be
mapped on the isotropic Heisenberg model in a random field). We assume the chain with periodic boundary conditions and $L$ sites. Furthermore, $t=1$ is used as the unit of energy.  

To probe transport response we evaluate the dynamical conductivity
$\sigma(\omega)$, 
\begin{equation}
\sigma(\omega) = \frac{1-{\rm e}^{-\omega/T}}{ \omega~ L} 
Re \int_0^\infty dt {\rm e}^{i \omega t}  \langle j(t) j\rangle~, \label{sig}
\end{equation}
with the current operator
\begin{equation}
j= i t \sum_i( c^\dagger_{i+1} c_i  - {\rm h.c.}) ~. \label{j}
\end{equation}
We adopt the view that possible conductor-insulator transition needs to be  
a manifestation of the character of MB quantum states\cite{ogan,zpp,mont} 
(hence not  directly related to other thermodynamic quantities). Therefore 
one may as well restrict the study to the regime $T \to \infty, \beta \to 0$,
where all MB states contribute with equal weight to $\sigma(\omega)$.  
In this limit, the relevant (and nontrivial) quantity is 
$\tilde \sigma(\omega)=T \sigma(\omega)$. 

$\tilde \sigma(\omega)$ is calculated by employing the microcanonical Lanczos method 
(MCLM)\cite{mclm,kara}, particularly suited for dynamical quantities
at elevated $T$.  In the following we present results for systems with $L=16 - 24$ sites
and for generic cases of half- and quarter-filling $n=N_e/L = 1/2, 1/4$, respectively. 
A sampling over $N_r \sim 100$ random $\epsilon_i$ configurations
is made to obtain the relevant average response.    
Finite size effects should not affect significantly our analysis in both, the energy 
and space domains. Concerning the space domain, we focus on disorder parameters $W$ 
for which the NI-electron localization length $\xi_0$ is much shorter than the size 
of the system, $\xi_0\ll L$. One may use the estimate (for $V=0$) $\xi_0 \sim 28.5/W^2$,\cite{schm} which for $W=2-4$ gives $\xi_0=7-1.8\ll L$. Moreover, $T=0$ DMRG calculations 
indicate that the effective $\xi$ is further reduced by the interaction $\Delta>0$.\cite{schm} 

The energy resolution of our spectra is much smaller than the 
average level spacing associated with the largest system size $L=24$ studied. 
Approximate eigenfunctions corresponding to $T \to \infty$ limit 
are converged in $M_1 \sim 2000$ Lanczos steps, providing the energy resolution 
of $\delta E \sim 0.004$ (for $L=24$). In the next step, $M_2\sim4000$ 
Lanczos iterations are used 
to evaluate $\tilde \sigma(\omega)$, leading to an estimation of the final frequency resolution 
$\delta \omega \sim 0.00$5. Since for the largest $L=24$ the studied sector contains $N_{st} \sim 2\;10^6$ MB states, the average level spacing $\Delta E \sim 10^{-5} \ll \delta \omega$, so the discreteness of the exact eigenspectrum due 
to finite system size plays practically no role in our results.

\begin{figure}[htb]
\includegraphics[width=0.8\linewidth]{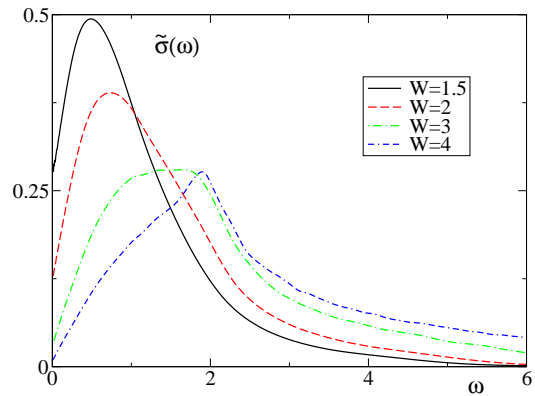}
\caption{(Color online) Dynamical high-$T$ conductivity $ \tilde \sigma(\omega)$ 
for $\Delta =0.5$ and different disorders $W=1.5,2,3,4$ evaluated for a half-filled system 
$n=1/2$ and $L=24$ sites.}
\label{fig1}
\end{figure}

In Fig.~1 we present typical high-$T$ spectra for $\tilde \sigma(\omega)$, showing 
different $W=1.5,2,3,4$ for fixed $\Delta =0.5$. Since we deal with a substantial 
disorder, $\tilde\sigma(\omega)$ are  essentially different from the 
weak-scattering Drude-like form. All curves in Fig.~1 reveal maxima at  $\omega_m >0$ 
moving up in frequency with increasing $W$, being a characteristic of a quasi-localized regime. 
At the same time, the optical sum rule is for $T\rightarrow\infty$ 
independent of $W$ and $\Delta$,\cite{kara}
\begin{equation}
\int_0^\infty  \tilde \sigma(\omega) d \omega= \frac{\pi}{2 L} \langle j^2 \rangle = \pi t^2 n(1-n), 
\label{sum}
\end{equation}
whereas the d.c. value $\tilde \sigma_0=\tilde \sigma(0)$, 
being the central quantity studied further on, shows a pronounced variation with $W$. 

\begin{figure}[htb]
\includegraphics[width=0.8\linewidth]{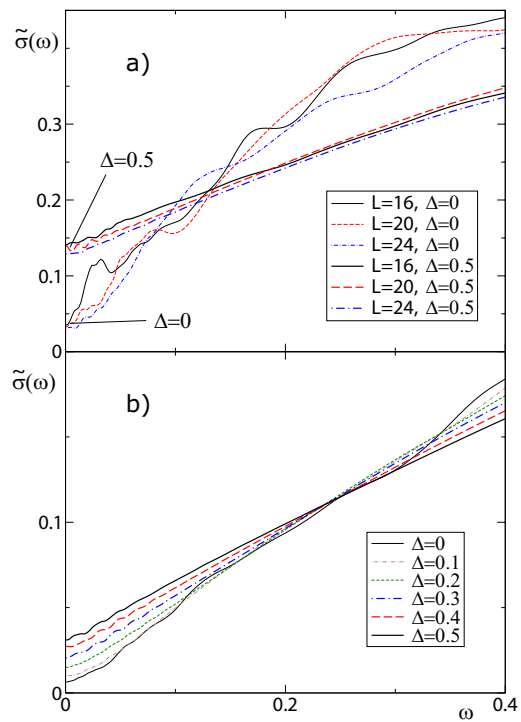}

\caption{(Color online) Low-$\omega$ part of $\tilde\sigma(\omega)$ for $n=1/2$: 
a) $\Delta=0, 0.5$, disorder $W=2$ and different $L=16,20,24$, 
b) $W=3$, $L=24$ and different $\Delta=0-0.5$.}
\label{fig2}
\end{figure}

In Figs.~2a,b the emphasis is given to the low-$\omega$ window, which is relevant 
for the extraction of the d.c. value $\tilde \sigma_0$. $W=2$ spectra for the 
$n=1/2$  case are smoothed with an $\omega$ dependent damping 
$\eta=\eta_0+(\eta_{\infty}-\eta_0) {\rm tanh}^2(\omega/\omega_0)$, 
$\eta_0=0.002\lesssim\delta \omega$, $\eta_{\infty}=0.02$, $\omega_0=0.2$. 
Such a damping, used hereafter, preserves the sensitivity for lowest $\omega\to0$
frequencies, while smoothing configuration-dependent fluctuations (most pronounced 
at $\Delta \to 0$) at higher $\omega$. 

As argued above for $\xi_0<L$ cases, Fig.~2a confirms the absence of any evident $L$ dependence (at least for $\Delta=0.5$). One can make an additional observation
that, unlike for $\Delta=0.5$, the fluctuations of $\tilde \sigma(\omega)$ 
even at low frequencies remain substantial for the NI case $\Delta=0$. One finds that these
fluctuations diminish with the increase of the sampling $N_r$ over random disorder 
configurations,  indicating that the repulsive interaction $\Delta>0$ suppresses the 
sensitivity to the
particular disorder configuration. In addition, $\Delta=0.5$ case in Fig.~2a reveals a 
remarkable linearity, $\tilde \sigma(\omega)\sim\tilde\sigma_0+\alpha|\omega|$,
being apparently generic\cite{kara} for all $\Delta>0$.

By varying $\Delta=0-0.5$, the role of the interaction on the MB localization is 
investigated in Fig.~2b. Results are given for fixed disorder $W=3$, $L=24$ 
and $n=1/2$. Again, the remarkable linearity $\tilde\sigma(\omega)$ at low-$\omega$ may be observed being very reproducible for $\Delta>0.1$ in spite of very small 
d.c. values $\tilde \sigma_0$ involved. Even more important, there is no signature 
of a presumable qualitative change in $\tilde \sigma(\omega)$ (at least for $\Delta>0.1$), 
which would point to the crossover for finite $\Delta$ from the MB localization, 
present at $\Delta=0$, to a conducting regime $\tilde \sigma_0 >0$.
   
Based on the same numerical analysis of $\tilde\sigma(\omega)$ used for Fig.~2a,b, 
in Fig.~3 we show the extracted d.c. values $\tilde\sigma_0$ vs. $\Delta$. To suppress 
the effects of the configuration-fluctuating component of 
$\tilde\sigma(\omega)$ as much as possible, $\tilde\sigma_0$ is evaluated from a linear fit of 
$\tilde\sigma(\omega)$ in the frequency interval $\omega<0.1$. $\tilde\sigma_0$ calculated 
in this way are given by symbols for $W=2,3,4$, respectively. For $\Delta<0.1$, 
dashed lines are used in Fig.~3 to interpolate $\tilde\sigma_0$ to the theoretical value 
$\tilde\sigma_0=0$ of the NI $\Delta\rightarrow0$ limit.

Results in Fig.~3 are central to this work. In spite of small values of $\tilde\sigma_0$, 
in particular for the $W=4$ case, the extracted $\tilde\sigma_0$ show very consistent behavior. 
Namely, it is quite evident that in the interval $W=2-4$ we do not find any signature of possible  
crossover in  the behavior $\tilde \sigma_0$ vs. $\Delta$, which could be interpreted as the 
onset of the MB localization for $\Delta<\Delta_c(W)$. In fact, the simplest 
dependence $\tilde\sigma_0\propto \Delta$ seems to represent well 
our results in the investigated regime $2 \leq W \leq 4$.

Since the possible MB localization at $\Delta>0$ should be more plausible at lower doping, 
where the condition $\xi<1/n$ is stronger, we investigate the quarter-filling 
case $n=1/4$ as well. Results for $\tilde \sigma_0$ vs. $\Delta$ are, however, even 
quantitatively similar to the $n=1/2$ case, although, as expected, somewhat 
smaller values of $\tilde \sigma_0$ are obtained. As presented in Fig.~4, it is instructive to follow the dependence of the d.c. value 
$\tilde\sigma_0$ as function of the disorder $W$. We investigate the $\Delta=0.5$ case 
for two fillings, $n=1/2,1/4$, and the data for $W=1.,1.5$ are included. It is evident from Fig.~4 that 
the dependence is exponential, i.e. $\tilde\sigma_0\sim a~{\rm exp}(-bW)$, 
with $b\approx1.7,2$ for $n =1/2, 1/4$, respectively.

\begin{figure}[htb]
\includegraphics[width=0.8\linewidth]{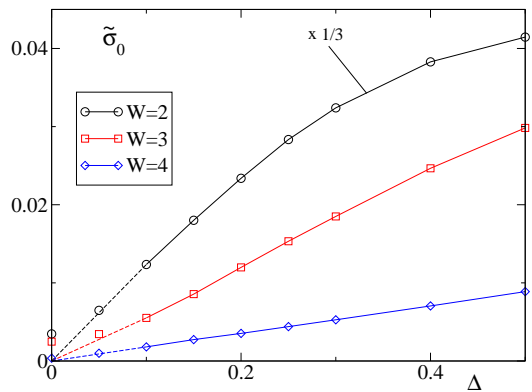}
\caption{(Color online) D.c. conductivity $\tilde\sigma_0$ vs. $\Delta$ for half-filling $n=1/2$ and 
different $W=2,3,4$. Dashed lines for $\Delta <0.1$ are interpolations
to the theoretically known value $\tilde \sigma_0(\Delta=0)=0$.}
\label{fig3}
\end{figure}
 
\begin{figure}[htb]
\includegraphics[width=0.8\linewidth]{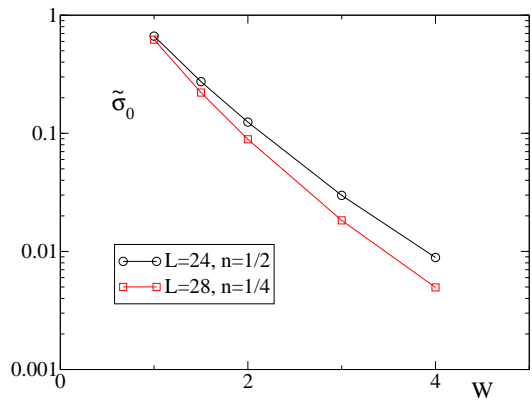}
\caption{(Color online) $ \tilde \sigma_0$ vs. $W$ for $\Delta =0.5$ and fillings $n = 1/2, 1/4$.}
\label{fig4}
\end{figure}

Fig.~4 gives also a clear limitation to our numerical approach in the regime $W\gg1$. 
Since the low-$\omega$ slope $\alpha$ of $\tilde \sigma(\omega)$ shows weaker 
dependence on $W$ ($\alpha\propto1/W$ from Fig.~1), a reliable evaluation of $\tilde\sigma_0$ 
requires a very high resolution $\delta \omega \ll 1$. The latter is determined in our MCLM 
method by $M_1, M_2$ Lanczos steps, but in the final stage also by the MB level density and 
the average MB level spacing $\Delta E \propto 1/N_{st}\propto {\rm exp}(-\zeta L)$. The reason for this is that 
macroscopic results for the transport become plausibly relevant (for $\Delta>0$ case) 
only if $\delta \omega \gg \Delta E$.

The above results show that a weak repulsive interaction in 1D disordered
tight-binding systems is capable of destroying the phenomena of MB localization. At
the present stage of investigations, we do not have a clear analytical or
phenomenological explanation for this property. However, we have so far discussed
cases by keeping the fermion density $n$ fixed. On the other hand, one can
investigate the behavior of the MB states by fixing the number of fermions $N_e$
while changing the system size $L$. Several such studies have been
reported,\cite{shep} suggesting that the two-particle localization length $\xi$ in
the presence of interaction becomes enhanced in comparison with the single particle
case. Because the context investigated has been rather different, we study here this effect
from the quantity directly relevant to the coherent charge transport, i.e. from the
charge stiffness $D$. For $T > 0$, $D$ is defined by\cite{znp}

\begin{equation}
D = \frac{\beta }{Z} \sum_n {\rm e}^{-\beta E_n} |\langle n|j|n  \rangle|^2. \label{d}
\end{equation}

\noindent As before, we are focused on the high-$T$ regime $\beta \to 0$, when all
the levels are probed, $Z=N_{st}$. In the absence of disorder and for finite $\Delta>0$, $\tilde D=T D$ remains finite because of the integrability of the
model.\cite{znp} With disorder switched on and $N_e$ fixed, one expects an
exponential suppression of $\tilde D$ with the increase of the system size $L$,
$\tilde D \propto {\rm exp}(-L/\xi)$. Furthermore, for the particular case of NI fermions $\xi$
should be independent of $N_e$, i.e., $\xi = \xi_0$, with $\xi_0$ denoting the
single fermion localization length.

Since for the localized phase (e.g. for $V>0$ and $T=0$) the stiffness $\tilde D$ is
distributed for different realizations of the disorder according to the log-normal
distribution,\cite{schm} we present in Fig.~5 the average $\bar D = {\rm exp}
(\langle {\rm ln} \tilde D\rangle)$ vs. $L$, with $N_e=2$, $\Delta=1$ and various
$W=0 - 4$. $\bar D$ is obtained by the full diagonalization (for $N_e=2$, $N_{st}
\propto  L^2$). It is evident from the figure that, for larger disorders, $W\geq 2$,
$\bar D$ decays exponentially with $L$, which is consistent with the MB localization
in the $n \to 0$ limit. On the other hand, it is remarkable that the interaction
$\Delta=1$ increases $\bar D$ at large $L$. It seems from Fig.~5 that a crossover
between the $\xi<\xi_0$ and $\xi>\xi_0$ behaviors occurs for $W\approx2$ and $L^*
\approx40$. For strong disorder $W=3,4$, the crossover appears already
at $L^* \sim 10$. These results suggest that the increase of $\xi$
due to interaction may be an argument for the elimination of the MB localization at
high $T$ for finite fermion densities $n=1/4,1/2$ discussed in this work.
 
\begin{figure}[htb]
\includegraphics[width=0.8\linewidth]{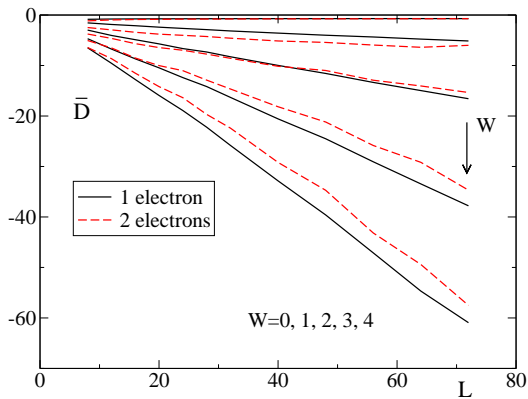}
\caption{(Color online) Normalized charge stiffness $\tilde D/N_e$ vs. $L$ 
for $N_e =1,2$ fermions with the interaction $\Delta =1.0$, for various $W=0-4$.  }
\label{fig5}
\end{figure}

In conclusion, our numerical results reveal a steady and uniform
increase of $\tilde \sigma_0$ with the repulsive interaction $\Delta>0$ at high $T$ and at fixed disorder $W$, which  
does not support a possible MB localization at $\Delta>0$ in the considered regime.
In order to put our results in broader context, let us comment the relation to other works. Authors advocating a finite-$T$ insulator-conductor transition\cite{bask}
give an estimate for the transition temperature 
$T^* \propto 1/({\cal N} \xi \lambda {\rm ln} \lambda)$ where ${\cal N}$ is the 
(single-particle) density of states and $\lambda$ a characteristic matrix
element for the electron-hole pair creation. For high $T \gg 1$  
and weak interaction $\Delta<1$ one may translate the estimate\cite{ogan} to a critical 
$\Delta^* \propto \lambda^* \propto 1/({\cal N}\xi_0)$. 
Hence, for larger $W>2$ we are in the regime where at least some crossover 
should be observed for $\Delta \sim \Delta^* >0$. Still, we do not observe 
any clear sign of the latter, at least it is not evident enough.

On the other hand, some recent numerical studies using different criteria, 
seem to point to the possible conductor-insulator transition and 
the MB localization at high $T$ at much  larger  $W$, 
essentially within the same model with typically fixed 
interaction $\Delta \sim 1$.\cite{mont,berk,pal} 
Translating the definitions of disorder $W$, their estimates for the onset of 
localization would be $W>W^* \sim 6 - 10$,
consistent with the observed qualitative change of the level statistics.\cite{ogan} It should be observed that such cases correspond to extreme disorder, which would require within our (or an analogous) approach the observation
(see Fig.~4) of $\tilde \sigma_0 < 10^{-4}$. The corresponding resolution 
$\delta \omega<10^{-4}$ and large MB density-of-states leading to $
\Delta E \propto LW / N_{st} <\delta  \omega $ may be in principle obtained e.g.
by the full diagonalization for  large enough $L$. However, the latter is not reachable by 
up-to-date numerical methods. Hence, we cannot exclude such a
scenario for the onset of the MB localization at $W>W^*$, but on the other hand, 
such extreme disorder would also put limits to its theoretical as well as 
experimental verification and relevance.

We authors acknowledge helpful discussions with X. Zotos.

\end{document}